\documentclass{article}
\usepackage[T1]{fontenc} 
\usepackage[utf8]{inputenc} 
\usepackage{ismir,amsmath,cite,url}
\usepackage{graphicx}
\usepackage{color}
\def\lbd{}	

\usepackage{lineno}

\title{DawDreamer: Bridging the Gap Between Digital Audio Workstations and Python Interfaces}

\oneauthor
{David Braun} {Center for Computer Research in Music and Acoustics \\ Stanford University \\ {\tt braun@ccrma.stanford.edu}}

\def\authorname{D. Braun}

\begin{document}

\maketitle
\begin{abstract}
Audio production techniques which previously only existed in GUI-constrained digital audio workstations, live-coding environments, or C++ APIs are now accessible with our new Python module called DawDreamer. DawDreamer therefore bridges the gap between real sound engineers and coders imitating them with offline batch-processing. Like contemporary modules in this domain, DawDreamer can create directed acyclic graphs of audio processors such as VSTs which generate or manipulate audio streams.
DawDreamer can also dynamically compile and execute code from Faust, a powerful signal processing language which can be deployed to many platforms and microcontrollers.
We discuss DawDreamer's unique features in detail and potential applications across music information retrieval including source separation, transcription, and audio effect parameter inference. We provide fully cross-platform PyPI installers, a Linux Dockerfile, and an example Jupyter notebook.\footnote{\url{https://github.com/DBraun/DawDreamer}}
\end{abstract}

\section{Introduction}\label{sec:introduction}

A digital audio workstation (DAW) is a software system which integrates most music production tasks including composing, recording, editing, adjusting effects, and exporting to audio files. An audio engineer typically uses a mouse and keyboard or expensive mixing console to carry out these tasks, making it difficult to explore efficiently the large action space of effects and their parameters. Moreover, some digital instruments and effects are platform specific, such as Audio Units on macOS or LV2 plug-ins on Linux. The ideal batch-processing audio framework with relevance to machine learning should both overcome the hurdles of mouse-and-keyboard interfaces and unify instruments and effects across all platforms.

One project in this domain is RenderMan \cite{leon_fedden_2017_1079885}, a Python module which served as the starting codebase for DawDreamer. RenderMan uses the JUCE\cite{JUCE} framework for rendering audio from VST\footnote{VST is short for Virtual Studio Technology, an audio plug-in software interface licensed by Steinberg Media Technologies.} instruments. RenderMan played a crucial role in research on software synthesizer presets \cite{8323327,DBLP:journals/corr/abs-1907-00971,DBLP:journals/corr/abs-2104-03876} and massive audio generation \cite{manilow2019cutting}, but its development has been slow to branch into other aspects of music production such as bussing. Bussing is the summation of audio tracks as an intermediate step in some mixing procedure.
Other researchers tried RenderMan but transitioned to a Max/MSP method after encountering audio artifacts \cite{Sarroff2020BLINDAR}.

FluidSynth \cite{fluidsynth2011} is a sample-based synthesizer engine with command-line support, but its reliance on SoundFount samples limits broader applications. Pedalboard is a new project with similarities to RenderMan and DawDreamer \cite{Pedalboard}. It has a promising future but currently lacks support for Faust, parameter automation, efficient time-stretching and pitch-bending, and generalized bussing (audio processor graph building). 

\section{Features}\label{sec:features}

DawDreamer aims to address the limitations of other tools and expand the capabilities of Python interfaces which emulate DAWs. Users can compose graphs of audio processors and record multiple processors at once in a single forward-pass. Therefore one pass can efficiently produce mixed and unmixed audio tracks, which is ideal for machine learning pipelines. Graphs can be reused, and processors' settings can be adjusted for subsequent passes. Parameter automation, which is the automatic changing of parameters over time, can be accomplished by specifying control signals as \texttt{numpy} arrays.

DawDreamer introduces some audio processors not available in other packages. In the following sections, we will describe the support for (1) arbitrary VST instruments and effects, (2) Faust code, (3) time-stretching and pitch-warping.

\subsection{Virtual Sound Technology}\label{subsec:vsts}

Like RenderMan, DawDreamer supports VST instruments, but it also supports VST effects. Furthermore, it  supports VST effects that take multiple inputs such as a sidechain compressor that attenuates the volume of one input according to the loudness of another.

\subsection{FAUST}\label{subsec:faust}

Faust (\textbf{F}unctional \textbf{AU}dio \textbf{ST}ream) is a programming language for real time signal processing \cite{faust2009}. Faust's built-in libraries include functions for reverbs, compressors, oscillators, filters, ambisonics, Yamaha DX7 emulation, and more.\footnote{\url{https://faustlibraries.grame.fr}}

DawDreamer uses the \texttt{libfaust} \cite{libfaust2013} backend to compile Faust code just-in-time. Elements in the Faust source code that would usually designate user interfaces such as sliders or toggles instead become parameters which can be automated according to \texttt{numpy} arrays.

This same coupling between Faust user interfaces and DawDreamer enables easy control of polyphonic Faust instruments \cite{Letz2017PolyphonyS}. A developer can write Faust code with a single voice of polyphony in mind and provide MIDI notes from Python or from a MIDI file. All of the voice allocation is done automatically.

The Faust examples in DawDreamer include a sidechain compressor, polyphonic wavetable synthesizer, and polyphonic sampler instrument. The synthesizer's wavetable and the sampler's sample can be specified with \texttt{numpy} arrays. The sampler example shows the simplicity of using MIDI-triggered ADSR envelopes and information to modulate the sample's pitch, volume and filter cutoff. One no longer needs to compose \texttt{numpy} functions to slice, fade, or filter short audio samples in order to emulate a basic sampler.

Beyond DawDreamer, Faust code can be compiled for Windows, Linux, macOS, Android, iOS, and many microcontrollers such as Teensy, SHARC, Bela, and most recently FPGAs.\footnote{\url{https://fast.grame.fr}} It can also be exported in many project formats and languages such as JUCE, Max, vcvrack, rust, julia, soul, C, C++, and more.\footnote{The Faust IDE (\url{https://faustide.grame.fr}) is the best way to get started with exporting Faust code.} Researchers would be wise to not restrict themselves to VST and LV2 audio plug-ins when Faust can be deployed so widely.

\subsection{Time-Stretching and Pitch-Warping}\label{subsec:rubberband}

DawDreamer borrows from a "warp marker" concept developed by the Ableton Live DAW \cite{AbletonLive} to provide an easy and efficient interface for time-stretching and pitch-warping audio. Each warp marker pairs a time in seconds and a position measured in beats. Ableton can generate and save warp markers to files with an \texttt{.asd} extension, which we reverse engineered.\footnote{A companion Python module is available: \url{https://github.com/DBraun/AbletonParsing}} Thus, DawDreamer can parse Ableton \texttt{.asd} files and use the Rubber Band Library \cite{Rubberband} to pitch-warp and time-stretch the associated audio without writing to the file system as an intermediate step like prior modules do \cite{audiomentations, pyrubberband}. The start/end markers and loop positions from the \texttt{.asd} file affect the audio's playback. One can also efficiently re-use the same clip at several places along a global timeline in DawDreamer's renderer.


\section{Potential Use Cases}\label{sec:potential_use_cases}

\subsection{Generative Mash-ups and Music Information Retrieval}\label{subsec:mashups}

Research on adversarial semi-supervised audio source separation would benefit from more ways to generate mixed and unmixed tracks with variations in timing and pitch \cite{DBLP:journals/corr/abs-1711-00048}. Therefore, we provide a Jupyter notebook\footnote{An automatically annotated example output can be seen at \url{https://youtu.be/HkK2ocYSUL0}} that tempo-matches and mixes a cappella and instrumental pairs according to an L2 distance combining their proximity in beats per minute and the musical circle of fifths.

A researcher of universal music source separation could use DawDreamer and generative music composition networks to create ground truth mixtures of tens of audio tracks rather than the common four (vocals, drums, bass, and other) \cite{musdb18-hq}. With adversarial learning, these generated mixtures could become increasingly realistic and helpful for source separation, transcription, lyrics alignment, instrument identification, cover identification, and more.

%
%
%

\subsection{Intelligent Music Production}

In the task of automatic audio mastering, DeepAFX achieved high quality results through gradient approximation of a fixed series of LV2 audio effects \cite{DeepAFX2021}. DeepAFx also succeeded at picking plug-in parameters to match a guitar pedal's distortion. In both cases, DawDreamer could learn the same mastering or compressor with Faust effects, but thanks to Faust, the effect could be deployed easily to more microcontrollers.

DawDreamer has potential applications in not only intelligent \textit{effects} but also intelligent signal \textit{generators}. Previous research on synthesizer parameter inference or exploration \cite{8323327,DBLP:journals/corr/abs-1907-00971,DBLP:journals/corr/abs-2104-03876,huang2014active, scurto2021designing} has been constrained by black-box compiled synthesizer code and plug-in formats, but DawDreamer can run arbitrary signal generators written with Faust. For example, the Slakh project\cite{manilow2019cutting} relied on presets and sample packs for the Native Instruments' plug-in Kontakt, but DawDreamer can pass audio samples to polyphonic Faust signal generator code, either of which could be learned via some algorithm.

\section{Conclusion}\label{sec:conclusion}

Much of music production is a series of actions taken inside a DAW environment\footnote{Perhaps Reinforcement Learning researchers can also begin to think of the DAW as an environment, just like an Atari video game.}, yet some ML researchers study musical audio as a raw series of numbers. To be fair, this domain-agnosticism helps models generalize to other domains, but it forfeits the helpful inductive biases from understanding music as the interaction of MIDI notes, sample packs, signal chains, effects, and parameter settings. Those building blocks and domain knowledge form a large part of the DNA of music. Researchers can now use DawDreamer as the physically unconstrained software engine that grows musical DNA into fully-realized audio data.


%
%
%

\section{Acknowledgments}

The author thanks Leon Fedden for starting RenderMan and making it open-source; Julius O. Smith III and Stéphane Letz for their support with Faust; Christian Steinmetz and Chris Donahue for their feedback on the manuscript.


\bibliography{ISMIR2021_lbd}

\begin{thebibliography}{10}
\providecommand{\url}[1]{#1}
\csname url@samestyle\endcsname
\providecommand{\newblock}{\relax}
\providecommand{\bibinfo}[2]{#2}
\providecommand{\BIBentrySTDinterwordspacing}{\spaceskip=0pt\relax}
\providecommand{\BIBentryALTinterwordstretchfactor}{4}
\providecommand{\BIBentryALTinterwordspacing}{\spaceskip=\fontdimen2\font plus
\BIBentryALTinterwordstretchfactor\fontdimen3\font minus
  \fontdimen4\font\relax}
\providecommand{\BIBforeignlanguage}[2]{{%
\expandafter\ifx\csname l@#1\endcsname\relax
\typeout{** WARNING: IEEEtran.bst: No hyphenation pattern has been}%
\typeout{** loaded for the language `#1'. Using the pattern for}%
\typeout{** the default language instead.}%
\else
\language=\csname l@#1\endcsname
\fi
#2}}
\providecommand{\BIBdecl}{\relax}
\BIBdecl

\bibitem{leon_fedden_2017_1079885}
\BIBentryALTinterwordspacing
L.~Fedden, ``{fedden/RenderMan: The v1.0.0 release for publication of paper},''
  2017. [Online]. Available: \url{https://doi.org/10.5281/zenodo.1079885}
\BIBentrySTDinterwordspacing

\bibitem{JUCE}
\BIBentryALTinterwordspacing
J.~Storer, ``{JUCE}: Jules’ utility class extensions,'' London, U.K., 2010.
  [Online]. Available: \url{https://www.juce.com/}
\BIBentrySTDinterwordspacing

\bibitem{8323327}
M.~J. Yee-King, L.~Fedden, and M.~d'Inverno, ``Automatic programming of {VST}
  sound synthesizers using deep networks and other techniques,'' \emph{IEEE
  Transactions on Emerging Topics in Computational Intelligence}, vol.~2,
  no.~2, pp. 150--159, 2018.

\bibitem{DBLP:journals/corr/abs-1907-00971}
\BIBentryALTinterwordspacing
P.~Esling, N.~Masuda, A.~Bardet, R.~Despres, and A.~Chemla{-}Romeu{-}Santos,
  ``Universal audio synthesizer control with normalizing flows,'' \emph{CoRR},
  vol. abs/1907.00971, 2019. [Online]. Available:
  \url{http://arxiv.org/abs/1907.00971}
\BIBentrySTDinterwordspacing

\bibitem{DBLP:journals/corr/abs-2104-03876}
\BIBentryALTinterwordspacing
C.~Mitcheltree and H.~Koike, ``{SerumRNN}: Step by step audio {VST} effect
  programming,'' \emph{CoRR}, vol. abs/2104.03876, 2021. [Online]. Available:
  \url{https://arxiv.org/abs/2104.03876}
\BIBentrySTDinterwordspacing

\bibitem{manilow2019cutting}
E.~Manilow, G.~Wichern, P.~Seetharaman, and J.~Le~Roux, ``Cutting music source
  separation some {Slakh}: A dataset to study the impact of training data
  quality and quantity,'' in \emph{Proc. IEEE Workshop on Applications of
  Signal Processing to Audio and Acoustics (WASPAA)}.\hskip 1em plus 0.5em
  minus 0.4em\relax IEEE, 2019.

\bibitem{Sarroff2020BLINDAR}
A.~M. Sarroff, ``Blind arbitrary reverb matching,'' 2020.

\bibitem{fluidsynth2011}
D.~Henningsson, ``{FluidSynth} real-time and thread safety challenges,'' in
  \emph{Proceedings of the 9th International Linux Audio Conference}, Maynooth
  University, Ireland, 2011, pp. 123--128.

\bibitem{Pedalboard}
\BIBentryALTinterwordspacing
{Spotify AB}, 2021. [Online]. Available:
  \url{https://github.com/spotify/pedalboard/}
\BIBentrySTDinterwordspacing

\bibitem{faust2009}
Y.~Orlarey, D.~Fober, and S.~Letz, \emph{FAUST: an Efficient Functional
  Approach to DSP Programming}, January 2009.

\bibitem{libfaust2013}
S.~Letz, D.~Fober, and Y.~Orlarey, ``Comment embarquer le compilateur faust
  dans vos applications ?'' May 2013.

\bibitem{Letz2017PolyphonyS}
S.~Letz and Y.~Orlarey, ``Polyphony, sample-accurate control and {MIDI} support
  for {FAUST} {DSP} using combinable architecture files,'' 2017.

\bibitem{AbletonLive}
\BIBentryALTinterwordspacing
 [Online]. Available: \url{https://www.ableton.com/}
\BIBentrySTDinterwordspacing

\bibitem{Rubberband}
\BIBentryALTinterwordspacing
[Online; accessed 12-September-2021]. [Online]. Available:
  \url{https://breakfastquay.com/rubberband/}
\BIBentrySTDinterwordspacing

\bibitem{audiomentations}
\BIBentryALTinterwordspacing
I.~Jordal, 2019, [accessed 12-September-2021]. [Online]. Available:
  \url{https://github.com/iver56/audiomentations}
\BIBentrySTDinterwordspacing

\bibitem{pyrubberband}
\BIBentryALTinterwordspacing
B.~McFee, ``pyrubberband,'' 2015. [Online]. Available:
  \url{https://github.com/bmcfee/pyrubberband}
\BIBentrySTDinterwordspacing

\bibitem{DBLP:journals/corr/abs-1711-00048}
\BIBentryALTinterwordspacing
D.~Stoller, S.~Ewert, and S.~Dixon, ``Adversarial semi-supervised audio source
  separation applied to singing voice extraction,'' \emph{CoRR}, vol.
  abs/1711.00048, 2017. [Online]. Available:
  \url{http://arxiv.org/abs/1711.00048}
\BIBentrySTDinterwordspacing

\bibitem{musdb18-hq}
\BIBentryALTinterwordspacing
Z.~Rafii, A.~Liutkus, F.-R. Stöter, S.~I. Mimilakis, and R.~Bittner,
  ``{MUSDB18-HQ} - an uncompressed version of {MUSDB18},'' 2019. [Online].
  Available: \url{https://doi.org/10.5281/zenodo.3338373}
\BIBentrySTDinterwordspacing

\bibitem{DeepAFX2021}
M.~A. Martínez~Ramírez, O.~Wang, P.~Smaragdis, and N.~J. Bryan,
  ``Differentiable signal processing with black-box audio effects,'' in
  \emph{ICASSP 2021 - 2021 IEEE International Conference on Acoustics, Speech
  and Signal Processing (ICASSP)}, 2021, pp. 66--70.

\bibitem{huang2014active}
C.-Z.~A. Huang, D.~Duvenaud, K.~C. Arnold, B.~Partridge, J.~W. Oberholtzer, and
  K.~Z. Gajos, ``Active learning of intuitive control knobs for synthesizers
  using gaussian processes,'' in \emph{Proceedings of the 19th international
  conference on Intelligent User Interfaces}, 2014, pp. 115--124.

\bibitem{scurto2021designing}
H.~Scurto, B.~V. Kerrebroeck, B.~Caramiaux, and F.~Bevilacqua, ``Designing deep
  reinforcement learning for human parameter exploration,'' \emph{ACM
  Transactions on Computer-Human Interaction (TOCHI)}, vol.~28, no.~1, pp.
  1--35, 2021.

\end{thebibliography}

%
%
%
%
%

\end{document}